\documentclass[a4paper,10pt]{article}
\usepackage{latexsym,amsmath,mathrsfs}

\def\lag{\mathscr L}

\title{\Huge  On Topological Defects and Cosmological Constant}
  \author{B. Raychaudhuri\thanks{E-mail: biplabphy@visva-bharati.ac.in},
  F. Rahaman\thanks{E-mail: rahaman@iucaa.ernet.in},
   M. Kalam\thanks{E-mail: mehedikalam@yahoo.co.in} \\
$^\ast$\small Department of Physics, Visva-Bharati, Santiniketan, West Bengal, 731235 India\\
 $^\dag$\small Department of Mathematics, Jadavpur University,
Kolkata - 700032, India
\\
$^\ddag$\small Department of Physics, Aliah University, Salt Lake, Kolkata -700091,  India,\\
 }

\begin{document}

  \maketitle
 \begin{abstract}
Einstein introduced Cosmological Constant in his field equations in an ad hoc manner. Cosmological constant plays the role of vacuum energy of the universe which is responsible for the accelerating expansion of the universe. To give theoretical support it remains an elusive goal to modern physicists. We  provide a prescription to obtain cosmological constant from the phase transitions of the early universe when topological defects, namely monopole might have existed.
\end{abstract}


The observational result that the universe is expanding at a rate which is increasing with time, rather than in a constant rate contrary to the established  belief was announced by two groups in 1999~\cite{riess, perl}.The observation was made on type Ia supernova. This result was also proved by WMAP survey~\cite{bennett, spergel}. The observation was further corroborated by WMAP data analysis~\cite{bennett}. The acceleration of expansion essentially means that the scale factor in the standard FRW cosmology possesses a positive second derivative. The discovery has the same classic status as the Hubble's observation of expansion of universe which had earlier been believed to be static. It is a well-known fact  in the history of physics how Einstein introduced cosmological constant in GR to balance the expansion of universe theoretically predicted by Lama\^itre and finally retracted calling the introduction the greatest blunder of his life. On the other hand at the time of finding the observational result of the accelerating expansion there were no theoretical prediction. It is expected that the acceleration is caused by the presence of certain kind of field whose effect is to put negative pressure, thus introducing a force similar to gravity but repulsive in nature. The notion of `dark energy' was introduced to account for the acceleration. The most suitable candidate for this mysterious force is the cosmological constant ($\Lambda$) conceived by Einstein. There are other scalar field models too  --  quintessence, K-essence, Chaplygin gas, tachyonic scalar and phantom energy.
 
Whatever be its form, it is expected that the observational result can be explained once some kind of repulsive gravitational effect is introduced in the theory. 

Notably, the  gravitational effect of global monopole is found to be repulsive in nature~\cite{bv,hl}. Thus one may expect that the global monopole and cosmological constants are connected through their common manifestation as the origin of repulsive gravity. Moreover, both cosmological constant and vacuum expectation value are connected while vacuum expectation value is connected to the topological defect.

All these points lead us to a simple conjecture. There must be a common connection among them, namely, the cosmological constant, the global monopole (topological defect) and the vacuum expectation value. In this paper, precisely this connection will be dealt upon -- although the present model of global monopole, which is the simplest one, will have to be modified. We show that the cosmological constant may have its origin at the time of symmetry breaking during the phase transition period of early universe.

In the systematic phase of early universe, there was no existence of topological defect. During the expansion of the early universe, after the Planck time, different phase transitions resulted in to the formation of various kind of much discussed topological defects like monopoles (point defects), cosmic strings (line defects) and domain walls (sheet defect)~\cite{kibble}. The topology of the vacuum manifold dictates the nature of these topological defects. These topological defects appeared due to breakdown of local or global gauge symmetries.

Global monopole is a heavy object formed as a result of gauge-symmetry breaking in the phase transition of a self-coupling iso-scalar triplet $\phi^a$ system. They are similar to elementary particles as major part of their energy remains concentrated in a small region near the monopole core.

The following  Lagrangian is the simplest model describing a
global
 monopole~\cite{vilenkinshellard} and has been discussed by
 many authors.
\begin{equation}
\lag = \frac{1}{2} \partial_\mu \phi^a \partial^\mu \phi^a -
\frac{1}{4}\lambda \left( \phi^a \phi^a - \eta^2 \right)^2.
\end{equation}

Here $\phi^a $ is a Higgs field scalar triplet ($a=1,2,3$). The components of Higgs field dictate the defects of various kinds -- singlet for domain walls, and doublet for cosmic string. $\eta$ is the vacuum expectation value and $\lambda$ is coupling constant. The Lagrangian is reminiscent of the $\phi^4$ field, now with a topological defect factor and a  rescaled coupling constant. The role of $\eta$ here is to produce a deficit angle in the metric of the space around the global monopole in its gravitational field~\cite{bv}. The deficit solid angle is proportional to the energy scale of the spontaneous symmetry breaking.

The topological defect term $\eta$ is also the self coupling strength of the triplet and
\begin{equation}
  \label{phia}
  \phi^a = \eta f(r) \frac{x^a}{r}
\end{equation}
where $x^ax^a=r^2$ and $f(r)$ is an arbitrary function of $r$. Note further that the potential term does not contain any mass term.

In this paper we slightly shift from the simplest model of the global monopole introducing a mass term in the potential. The Lagrangian now becomes

\begin{equation}
\lag =\frac{1}{2} \partial_\mu \phi^a \partial^\mu \phi^a -
\frac{1}{4}\lambda \left( \phi^a \phi^a - \eta^2 \right)^2 - m^2 \phi^a \phi^a
\end{equation}

The extra term $-m^2 \phi^a\phi^a$ introduced here reminds one to the familiar linear sigma model of field theory~\cite{weinbook}. The similar Lagrangian in the field theory obviously is devoid of the topological term and describes the pion-pion interaction. We shall show below that  the mass term accounts for
the appearance of cosmological constant. The Lagrangian is symmetric under rotation of $\phi^a$  and thus retains the symmetry breaking of the original global monopole model. It is therefore, proposed that the symmetry breaking of a linear sigma model with topological defect during the phase transition gives rise to the Cosmological Constant in the early Universe and hence relating the appearance of Cosmological Constant to the evolution of the early Universe.

To proceed on, we consider the general static metric with spherical symmetry,
\begin{equation}
ds^2 = e^{\nu(r)} dt^2 - e^{\mu(r)} dr^2 - r^2\left(d\theta^2 + \sin^2 \theta d \phi^2 \right)
\end{equation}
The field equation now reduces to
\begin{equation}
e^{-\mu} f'' + e^{-\mu} \left(\frac{2}{r} + \frac{\nu'}{2} - \frac{\mu'}{2}
 \right) f' - \frac{2f}{r^2} - \lambda \eta^2 f \left(f^2 -1 \right) + m^2 \eta f =0
\end{equation}
Following Barriola and Vilenkin~\cite{bv},  we consider the flat
space limit, i.e. $\nu=\mu=0$. The equation for $f(r)$ reduces to
\begin{equation}
f'' + \frac{2f'}{r} - \frac{2f}{r^2} - \lambda \eta^2 f \left(f^2 -1 \right) + m^2 \eta f =0
\end{equation}
We consider a simple power series form for $f(r)$ as the trial
solution
\begin{equation}
f(r) = \Sigma  a_n r^{-n}
\end{equation}
Putting this solution in the equation and equating the co-efficients of
 linearly independent terms to zero, we find
 \begin{equation}
 \begin{array}{l}
 a_0 = \left( 1 + \dfrac{m^2}{\lambda \eta} \right)^{1/2}\\
a_1 = 0\\
a_2 = \dfrac{2 a_0}{m^2 \eta +\lambda \eta^2 - 3 \lambda \eta^2 a^2_0}\\
a_3=0\\
a_4 = \dfrac{3 \lambda \eta^2 a_0 a_2^2}{m^2 \eta - 2 \lambda \eta^2 a_0^2 - \lambda \eta^2 (a_0^2 -1)}
 \end{array}
 \end{equation}
and so on.

Thus, $f(r)$ can be written as
\begin{equation}
f(r) = a_0 + \dfrac{a_2}{r^2} + \dfrac{a_4}{r^4} + {\mathscr O}\left( \dfrac{1}{r^6} \right)
\end{equation}
Note that, $f(r)$ is constant asymptotically and reduces to $a_0$. Also note that for $m=0$, $a_0 \longrightarrow 1$.

The energy momentum tensor is related to the Lagrangian by the
relation
\begin{equation}
T^{\mu \nu} = \dfrac{\partial \lag}{\partial (\partial_\mu \phi^a)} \partial^\nu \phi^a - g^{\mu\nu} \lag
\end{equation}
The components are given by
\begin{equation}
\begin{array}{l}
T^t_t = \frac{1}{2} e^{-\mu} \dfrac{\eta^2{f'}^2}{2} + \dfrac{\eta^2 f^2}{r^2}
 + \dfrac{1}{4} \lambda \eta^4 \left( f^2 -1 \right)^2 + \dfrac{1}{2} m^2 f^2 \eta^2\\
T^r_r = - e^{-\mu} \dfrac{\eta^2 {f'}^2}{2} + \dfrac{\eta^2 f^2}{r^2} + \frac{1}{4}
\lambda \eta^4 \left( f^2 -1 \right)^2 + \frac{1}{2}m^2 f^2 \eta^2 \\
T^\theta_\theta = T^\phi_\phi = \frac{1}{2} e^{-\mu} \eta^2 {f'}^2 + \frac{1}{2}m^2
f^2 \eta^2
\end{array}
\end{equation}
Putting $f(r) \simeq a_0$, we obtain ( neglecting the higher
orders of $r^{-n}$,
 meaning we are near the core)
\begin{equation}
\begin{array}{l}
T^t_t = \dfrac{\eta^2 a_0^2}{r^2} + \frac{1}{4} \lambda \eta^4 \left( a_0^2 -1
 \right)^2 + \frac{1}{2} m^2 a_0^2 \eta^2 = \dfrac{\eta^2 a_0^2}{r^2} + A \\
T^r_r = \dfrac{\eta^2 a_0^2}{r^2} +A \\
T^\theta_\theta = T^\phi_\phi =A
\end{array}
\end{equation}
where
\begin{equation}
 A = \frac{1}{4} \lambda \eta^4 \left( a_0^2 -1 \right)^2 + \frac{1}{2} m^2 a_0^2 \eta^2 .
\end{equation}
 Note that for massless field , $m=0$ implying $A=0$.

The new field equations now become
\begin{equation}\label{newfieldeqn}
\begin{array}{l}
e^{-\mu} \left( \dfrac{\mu'}{r} + \dfrac{1}{r^2} \right) + \dfrac{1}{r^2} = 8 \pi
 G \dfrac{a_0^2\eta^2}{r^2} + 8 \pi G A \\
 e^{-\mu} \left( \dfrac{\nu'}{r} - \dfrac{1}{r^2} \right)- \dfrac{1}{r^2} = 8 \pi G
\dfrac{a_0^2 \eta^2}{r^2} + 8 \pi G A \\
e^{-\mu} \left( \dfrac{\nu''}{2} - \dfrac{\mu'\nu'}{4} + \dfrac{\nu'^2}{4} +
\dfrac{\nu'-\mu'}{2r} \right) = 8 \pi G A
\end{array}
\end{equation}
The solution for the field equation is
\begin{equation}\label{newfieldsoln}
e^{-\mu} = e^\nu = 1- 8 \pi G \eta^2 a_0^2 - \dfrac{C}{r} - \dfrac{8 \pi G}{3} A r^2
\end{equation}
where $C$ is an integration constant and can be identified to the
mass of the monopole core.

It is clear that according to the Eqs.~\eqref{newfieldeqn} and \eqref{newfieldsoln} the term $8\pi GA$ plays the role 
 of cosmological constant~\cite{farookbul}. One is now prompted to conclude that 
 the spontaneous symmetry breaking of a Lagrangian of a linear sigma field
 with a scalar triplet from $O(3)$ to $U(1)$ gives rise to the Cosmological
  Constant which was first introduced by Einstein in an ad-hoc manner.
  Although Zeldovich ~\cite{zel} gave a unique quantum mechanical explanation
  for it, the present investigation shows that it emerged during the evolution of the Universe due
    to a spontaneous symmetry breaking.

The cosmological constant thus can be written as
\begin{equation}
 \Lambda = 8 \pi G A = 2 \pi G \lambda \eta^4 \left(a_0^2 -1 \right) + 4 \pi G m^2 a_0^2 \eta^2.
\end{equation}
Putting the value of $a_0^2= 1 + m/\lambda \eta^2$ we obtain
\begin{equation}
\label{Lambda}
 \Lambda = 8 \pi G \eta^2 \left( \frac{3}{4} \dfrac{m^4}{\lambda \eta^2} + \dfrac{1}{2} m^2 \right).
\end{equation}
This expression clearly shows the Cosmological Constant $\Lambda$
and $8\pi GA$ are dimensionally equivalent.
 The dimensionless quantity $8 \pi G\eta^2 \simeq 10^{-6}$.
$\eta$ here is the vacuum expectation energy of symmetry breaking and
is given by $\eta \sim 10^{16}$~GeV.
 $\lambda$ is a dimensionless coupling constant. The present observational estimated
 value of $\Lambda$ is $2.036 \times 10^{-35} s^{-2}$~\cite{carmeli}. Clearly it is
 possible to monitor the free parameters $m$ and $\lambda$ to obtain the  value of $\Lambda$.
It is also well known that the cosmological constant related to the vacuum expectation energy~\cite{carroll,weinberg,peeblesratra}. The presence of $\eta$ in Eq.~\eqref{Lambda} corroborates this. Moreover, the recent Lambda-CDM model of the universe asserts that $\Lambda$ is positive. Precisely that has been predicted in Eqn.~\eqref{Lambda}.

The connection of cosmological constant, vacuum energy and repulsive gravity is quite familiar. Cosmological constant as candidate for dark energy model may be responsible for the increasing rate of expansion universe. On the other side, the relation of global monopole and repulsive gravity is well-known through the spontaneous symmetry breaking phase of early universe. Vacuum expectation energy plays an important role here. The possible conjunction between these two theories is the repulsive gravity. This paper shows precisely this. 

\vspace{.5cm}

\textbf{Acknowledgment: } The authors would like to thank the Inter-University
 Centre for Astronomy and Astrophysics (IUCAA), Pune, India, for hospitality during  visits under the Associateship programme where a part of the work has been done.

\end{document}